\documentclass{egpubl}
\Paper
\usepackage[T1]{fontenc}
\usepackage{dfadobe}  
\usepackage{cite}
\BibtexOrBiblatex
\electronicVersion
\PrintedOrElectronic
\ifpdf \usepackage[pdftex]{graphicx} \pdfcompresslevel=9
\else \usepackage[dvips]{graphicx} \fi
\usepackage{egweblnk}

\graphicspath{{figures/}{images/}{imgs/}{Figures/}{resources/}}

\usepackage{appendix}

\usepackage{microtype}
\usepackage{amsfonts}
\usepackage{amsmath}
\usepackage{nicefrac}
\usepackage{array}
\usepackage{booktabs}

\usepackage[capitalize]{cleveref}
\crefname{paragraph}{Sec.}{Secs.}
\crefname{section}{Sec.}{Secs.}
\Crefname{section}{Section}{Sections}
\Crefname{table}{Table}{Tables}
\crefname{table}{Tab.}{Tabs.}
\Crefname{figure}{Figure}{Figures}
\crefname{figure}{Fig.}{Figs.}
\Crefname{chapter}{Chapter}{Chapters}
\crefname{chapter}{Ch.}{Chs.}

\usepackage{multicol, multirow}
\usepackage{xcolor}
\usepackage{colortbl}
\definecolor{1st}{RGB}{102,194,164} %
\definecolor{2nd}{RGB}{178,226,226} %
\definecolor{3rd}{RGB}{237,248,251} %
\definecolor{1stText}{RGB}{57,146,116} %
\newcommand\tstrut{\rule{0pt}{2.4ex}}
\newcommand\bstrut{\rule[-1.0ex]{0pt}{0pt}}

\title[Bag of Tricks for Efficient INPC]{A Bag of Tricks for Efficient Implicit Neural Point Clouds}
\author[F. Hahlbohm et al.]
{\parbox{\textwidth}{\centering%
    Florian Hahlbohm$^{1}$\hspace{2pt}\orcid{0009-0004-8710-1433}\hspace{10pt}
    Linus Franke$^{2}$\hspace{2pt}\orcid{0000-0001-8180-0963}\hspace{10pt}
    Leon Overkämping$^{1}$\hspace{2pt}\orcid{0009-0001-7756-6702}\hspace{10pt}
    Paula Wespe$^{1}$\hspace{2pt}\orcid{0009-0007-2477-2725}\\
    Susana Castillo$^{1}$\hspace{2pt}\orcid{0000-0003-1245-4758}\hspace{10pt}
    Martin Eisemann$^{1}$\hspace{2pt}\orcid{0000-0002-8673-4405}\hspace{10pt}
    Marcus Magnor$^{1,3}$\hspace{2pt}\orcid{0000-0003-0579-480X}
}\\
\parbox{\textwidth}{\centering%
$^1$Computer Graphics Lab, TU Braunschweig, Germany\hspace{7pt} \texttt{\{hahlbohm,castillo,eisemann,magnor\}@cg.cs.tu-bs.de}\\
$^2$Visual Computing Erlangen, FAU Erlangen-Nürnberg, Germany\hspace{7pt}\texttt{\{firstname.lastname\}@fau.de}\\
$^3$University of New Mexico, USA\\
\url{https://fhahlbohm.github.io/inpc_v2/}
}
}
\begin{document}

\teaser{
 \includegraphics[width=\linewidth, keepaspectratio]{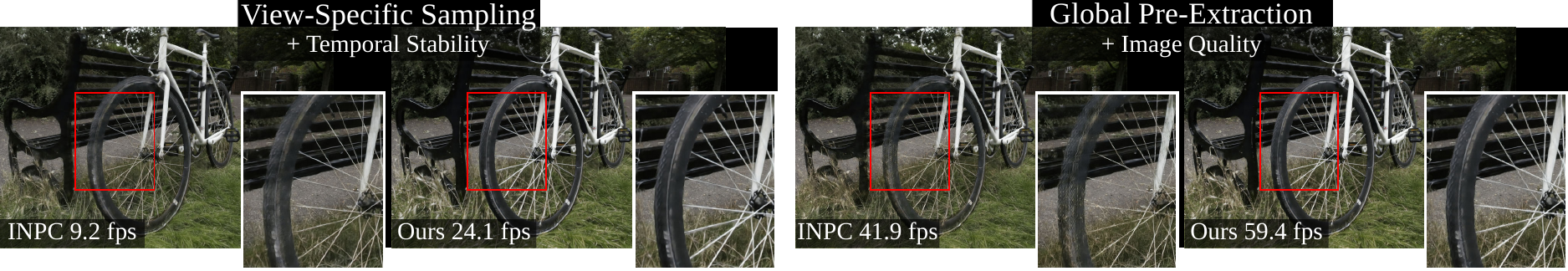}
 \centering
  \caption{Applying our bag to tricks to the original INPC~\cite{hahlbohm2024inpc} pipeline doubles rendering speed, reduces training time and memory usage, and improves image quality, enhancing its practical usability. We show our results alongside those of the original INPC implementation with the frame rates inset.
}
\label{fig:teaser}
}

\maketitle

\begin{abstract}
Implicit Neural Point Cloud (INPC) is a recent hybrid representation that combines the expressiveness of neural fields with the efficiency of point-based rendering, achieving state-of-the-art image quality in novel view synthesis.
However, as with other high-quality approaches that query neural networks during rendering, the practical usability of INPC is limited by comparatively slow rendering.
In this work, we present a collection of optimizations that significantly improve both the training and inference performance of INPC without sacrificing visual fidelity.
The most significant modifications are an improved rasterizer implementation, more effective sampling techniques, and the incorporation of pre-training for the convolutional neural network used for hole-filling.
Furthermore, we demonstrate that points can be modeled as small Gaussians during inference to further improve quality in extrapolated, e.g., close-up views of the scene.
We design our implementations to be broadly applicable beyond INPC and systematically evaluate each modification in a series of experiments.
Our optimized INPC pipeline achieves up to 25\% faster training, 2\texttimes\ faster rendering, and 20\% reduced VRAM usage paired with slight image quality improvements.
\end{abstract}

\section{Introduction}

The field of novel view synthesis, i.e., generating unseen views of a scene given a set of input images, has made rapid progress in recent years making immersive reconstructions of our world more realistic and feasible than ever.
Two recent breakthroughs, volumetric implicit fields with Neural Radiance Field (NeRF)~\cite{mildenhall2020nerf}, and point-based differentiable rendering via 3D Gaussian Splatting (3DGS)~\cite{kerbl3Dgaussians} have proven to be great representations for novel view synthesis.
Recently, Hahlbohm et al.~\cite{hahlbohm2024inpc} introduced the implicit neural point cloud (INPC), a hybrid representation combining the advantages of both to achieve state-of-the-art novel view synthesis results in terms of quality.
They reconstruct the scene volumetrically in a density octree (similar to NeRFs); however, they use conventional neural point-cloud rendering with small splats by extracting an explicit point representation from the octree.

Similar to other high-quality algorithms~\cite{barron2023ICCV}, INPC sacrifices rendering and training speed to achieve its high image quality, achieving rendering rates of around 3-10 frames per second, making it barely interactive.
This limits its usefulness in practice when compared to recent explicit particle-based representations such as 3DGS~\cite{kerbl3Dgaussians}, which not only render but also train much faster.

In this paper, we present a careful combination of improvements that aim at reducing computational demands for both the training and inference of INPC~\cite{hahlbohm2024inpc}.
Essential to our bag of tricks are the following extensions and modifications to the original algorithm:
\begin{itemize}
\item More effective extraction techniques, specifically temporally accumulated point clouds and well-distributed global point clouds, leading to major speedups and$\slash$or improved quality.
\item A significantly faster, more memory efficient, and flexible point rasterizer that also supports rendering neural points as small Gaussians during inference to improve generalization and reduce aliasing.
\item Modifications to the post-processing pipeline to allow for effective pre-training of the hole-filling CNN, which then makes optimization on unseen scenes converge faster and more robust.
\end{itemize}
We carefully evaluate these modifications and multiple other small improvements in a series of experiments to ensure they do not simply trade quality for speed. %
Applying our bag of tricks reduces training time by up to 25\% and speeds up rendering by up to 2\texttimes, while also significantly reducing VRAM requirements throughout the whole process.

As such, our contribution is twofold: (1) we propose improved rendering and filtering techniques to accelerate the neural implicit point cloud reconstruction and training; and (2) we provide refined approaches for more practical inference by introducing a more efficient view-specific temporal sampling technique, an algorithm that leads to more suitable point distributions when extracting large view-independent point clouds, and a specialized distance-agnostic splatting method using small isotropic Gaussians.

\section{Preliminaries}\label{sec:preliminaries}

Recent radiance field rendering techniques include volumetric neural fields represented as MLPs or feature grids~\cite{yu2021plenoxels, mueller2022instant, Chen2022tensorf, barron2022mipnerf360, barron2023ICCV, duckworth2023smerf, sun2025svraster}, point-based approaches~\cite{aliev2020npbg, ruckert2022adop, kopanas2021perviewopt, hahlbohm2023plenopticpoints, franke2023vet, franke2024trips, govindarajan2025radfoam}, and other particle-based methods extending~\cite{yu2024mip, radl2024stopthepop, kerbl2024hierarchical, kheradmand243dgsmcmc, taming3dgs} or building on top of 3DGS~\cite{kerbl3Dgaussians} in performance, quality, or capabilities~\cite{huang20242d, yu2024gof, moenne2024gaussiantracer, mai2024ever, hahlbohm2025htgs, franke2025vrsplatting, wu20253dgut, held2025convexsplatting, von2025linprim}.
A standard point-based neural rendering pipeline for novel view synthesis~\cite{aliev2020npbg} uses an explicit point cloud to represent the geometry of the scene.
Each point is defined by its 3D position $p_i\in\mathbb{R}^3$ and a feature vector $\omega_i\in\mathbb{R}^{n_\omega}$ that stores appearance information -- for a simple RGB color $n_\omega$ would be 3.
The point cloud can then be rendered to a 2D feature image $F\in\mathbb{R}^{H \times W \times n_\omega}$ using fast differentiable rasterization.
Due to the discrete nature of point clouds, the feature images may contain holes, which is typically solved by feeding it into a convolutional neural network (CNN) with a U-Net architecture~\cite{ronneberger2015unet}, to obtain the final RGB image $I\in\mathbb{R}^{H \times W \times 3}$.
Given a set of $N$ input photographs $I^{GT}_{1..N}$ and the corresponding camera parameters estimated by, e.g., COLMAP~\cite{schoenberger2016colmap1}, the point attributes and CNN weights can be optimized using gradient descent to enable novel view synthesis.
While existing works show this general architecture produces good quality in combination with high frame rates~\cite{ruckert2022adop, franke2024trips}, a major problem is the reliance on either a dense initial point cloud created by, e.g., slow multi-view stereo methods, or the application of heuristic techniques to densify a sparse initial point cloud during optimization.

In INPC~\cite{hahlbohm2024inpc}, Hahlbohm et al.\ argue that this reliance is a key reason why algorithms employing volumetric representations~\cite{barron2022mipnerf360, barron2023ICCV} achieve higher quality reconstruction and view synthesis results.
To address this, they propose a hybrid representation, where the combination of a probability field ($\mathcal{P}$) for point positions and a neural field ($\mathcal{A}$) for appearance features is used to represent a point cloud implicitly.

\paragraph{Implicit Point Cloud.}
The modeling of $\mathcal{P}$ is tackled by chopping the scene into a sparse grid of cubic voxels.
Beginning with a regular $128^3$ grid, they subdivide these voxels during optimization to capture intricate details. 

INPC uses a large multi-resolution hash grid~\cite{mueller2022instant} to encode point appearance at any 3D location.
To get these features at 3D point $p_i$, $\mathcal{A}$ is queried to produce a scalar opacity $o_i\in[0, 1]$ and Spherical Harmonics coefficients up to degree two $\hat{f}_i\in\mathbb{R}^{4 \times 9}$, resulting in four view-dependent features $f_i\in\mathbb{R}^{4}$ for a specific view direction.

\paragraph{Point Cloud Sampling.}
To extract explicit point clouds from $\mathcal{P}$, INPC uses different sampling strategies.
For viewpoint-specific sampling during training and inference, they use a reweighting scheme based on visibility, distance, and voxel size to achieve an effective sample distribution and query it uniformly for points.
Importantly, the viewpoint-specific reweighting causes the 2D feature images to be well-behaved in terms of the distribution of points on the image plane.
During inference, $\mathcal{P}$ can also be sampled globally, i.e., without a specific viewpoint in mind, to speed up rendering by avoiding per-frame point cloud sampling.
This is also called \textit{pre-extraction}.
Using these sampling techniques, an arbitrarily large set of point positions can be extracted from $\mathcal{P}$, which is then fed into $\mathcal{A}$ to obtain appearance features as described above ultimately yielding an explicit point cloud ready to be rendered.

\paragraph{Rendering Feature Maps.}
To obtain a 2D feature image $\mathcal{F}$ given the extracted point cloud, a differentiable point rasterizer with a bilinear splatting formulation is used.
The implementation of Hahlbohm et al.\ proceeds by creating a list of all created fragments and uses single radix sort with 64-bit keys combining pixel index and depth values to obtain a depth-sorted list of fragments for each pixel.
For each pixel in $\mathcal{F}$, all $K$ fragments are then blended together using standard alpha blending and composited with the output of a background model.

\paragraph{Post-Processing.}
Similar to other point-based neural rendering approaches, the feature image produced by the rasterizer may contain holes and also needs to be decoded as each pixel contains a multispectral feature.
Hahlbohm et al.\ use a large CNN $\mathcal{C}$, including a residual block based on Fast Fourier Convolution~\cite{chi2020fast} to reconstruct the completed image.
To increase robustness, they also employ the exposure and camera response modules from the differentiable tonemapper $\mathcal{T}$ proposed by Rückert et al.~\cite{ruckert2022adop} to map the high dynamic range output of the last block of the CNN
to sRGB space.

\paragraph{A Note on Sample Count.}
In their paper, Hahlbohm et al.~\cite{hahlbohm2024inpc} present three different configurations for the number of points sampled during each training iteration -- $2^{25}\approx33$M (their default), $2^{24}\approx16$M, and $2^{23}\approx8$M.
While a higher sample count leads to better image quality, it also significantly increases the VRAM requirements.
As a result, their larger models can not be trained on commodity hardware, i.e., GPUs with 24 GB VRAM or less.
A major motivation of our work is to make INPC more broadly applicable by improving its efficiency.
Therefore, we limit our evaluation to the configuration using 8M samples unless noted otherwise.
However, we are confident that the more expensive configurations would benefit equally from applying the presented tricks.

\section{Bag of Tricks}\label{sec:tricks}

In the following, we will describe the changes we made to INPC~\cite{hahlbohm2024inpc}, which are the result of a systematical analysis of the original training and inference pipeline (see \cref{sec:preliminaries}) and address performance bottlenecks regarding speed, quality, as well as memory usage.
We emphasize that this naturally causes our bag of tricks to include technical contributions, i.e., changes to the underlying algorithms, but also mere engineering contributions that have limited technical novelty outside of INPC.

\subsection{Improved Point Cloud Sampling}\label{ssec:sampling_tricks}

Hahlbohm et al.~\cite{hahlbohm2024inpc} present a novel approach for sampling points directly in 3D instead of along per-pixel rays~\cite{mildenhall2020nerf} or storing a persistent, optimizable set of points~\cite{aliev2020npbg, ruckert2022adop, kerbl3Dgaussians}.
We observe that, while their approach is highly optimized toward maximizing quality of single-frame renderings, it can be improved significantly to achieve higher frame rates and better temporal stability as well as better rendering quality in the global pre-extraction case.

\paragraph{Fast and Stable View-Specific Sampling during Inference.}
During inference, Hahlbohm et al.\ sample four view-specific point clouds for each frame, which improves quality of the renderings.
However, this also introduces a performance bottleneck as sampling $\mathcal{P}$ and $\mathcal{A}$ takes around 16 milliseconds per eight million points.
To address this problem, we propose to reuse the point clouds from previous frames.
As shown in \cref{fig:teaser} (left) and our experiments, our ring buffer approach significantly speeds up the rendering with barely any perceivable difference in quality.
An added benefit of this change is that it naturally improves temporal stability for INPC, an issue that point-based neural rendering methods commonly suffer from~\cite{ruckert2022adop, franke2024trips, hahlbohm2024inpc}.
While testing, we found that when the view-specific PDF contains a few very large weights and many small ones, too few points are created by the optimized inference sampling strategy of INPC.
Therefore, we modify the calculation of the per-voxel sample counts so that every visible voxel is at least sampled once.

\paragraph{Improved Global Pre-Extraction.}
To avoid repeated sampling of point clouds during inference, Hahlbohm et al.~\cite{hahlbohm2024inpc} propose a method for sampling a global, i.e., viewpoint-independent point cloud during inference.
While this speeds up rendering by up to 5\texttimes, it also reduces image quality by a sizable amount, which we find to be caused by too many points being placed in the background.
To ameliorate this, we devise an improved sampling strategy that better reflects how the model was trained and therefore leads to a more suitable distribution of points.
In contrast to Hahlbohm et al., we first compute a training view-agnostic PDF by first computing the maximum weight assigned to each voxel during view-specific sampling across all training views.
Similar to the original approach, we then sample the PDF to obtain a list of voxels, count the number of samples per voxel, and use per-voxel Halton sampling~\cite{halton1964} to place the final points.
We also find that with our method, we can significantly reduce the number of points from 67 to 33 million leading to significantly lower VRAM requirements.
The increased visual fidelity is exemplified in \cref{fig:teaser} (right).

\subsection{Efficient Feature Map Rendering}\label{ssec:rasterizer_tricks}

In their paper, Hahlbohm et al.\ identify sorting as the biggest bottleneck of the differentiable rasterizer.
As described in \cref{sec:preliminaries}, they create and subsequently sort four copies of each point, i.e., one for each pixel to which it could contribute.
While this results in per-pixel lists that can be processed collaboratively during blending (one warp blends one pixel), it also slows down sorting by 4\texttimes.
The complexity of the radix sort can be approximated as $\mathcal{O}(4kn)$, where $n$ is the number of points and $k$ is the key size.
As the used radix sort implementation from the CUB library launches one subroutine per 8 bits in the key and the key consists of a 32-bit depth and, e.g., a 21-bit pixel index for 1080p resolution, the complexity can be approximated as $\mathcal{O}(4 \cdot \lceil\nicefrac{(32+21)}{8}\rceil \cdot n) = \mathcal{O}(28n)$.

\paragraph{Tiled Rendering.}
To ameliorate the sorting cost, we adapt a tiled rendering approach similar to that used by Kerbl et al.~\cite{kerbl3Dgaussians} in 3DGS.
At a tile size of 8\texttimes8 pixels, it is easy to see that a 2\texttimes2 pixel-sized splat (as used in INPC) will on average contribute to $1.27$ tiles, resulting in a significant reduction in the number of copies being created.
Additionally, the key needs six fewer bits since it just holds the tile, not the pixel index.
This reduces the approximate complexity to $\mathcal{O}(1.27 \cdot \lceil\nicefrac{(32+15)}{8}\rceil \cdot n) = \mathcal{O}(7.62n)$.
As recently discussed by Schütz et al.~\cite{schuetz2025splatshop}, we can further reduce complexity by splitting the sorting into separate depth sorting of the points $\mathcal{O}(4n)$ and subsequent sorting of the created per-tile copies $\mathcal{O}(2.54n)$.
The final approximate complexity is therefore $\mathcal{O}(6.54n)$ a significant reduction compared to that of the original implementation.
An added benefit is that while we previously had to store $n$ $64$-bit keys as there is no $56$-bit data type, it is now sufficient to store $n$ 32-bit keys for the first and $1.27n$ 16-bit keys for the second sort, which reduces the required memory by 18\%.

\paragraph{Faster Background Model Evaluation.}
INPC uses an MLP to allow accurate modeling of low-frequency background, such as the sky, without placing sample points there.
Although the MLP is small, we observe that evaluating it for every pixel of the image to be rendered requires about one to two milliseconds depending on the image resolution.
To avoid this overhead, we distill the MLP into a spherical environment map represented as a 2D image $\mathcal{F}_{\text{bg}}\in\mathbb{R}^{1024 \times 2048 \times 4}$ after training that can be queried efficiently using bilinear interpolation.
We train $\mathcal{F}_{\text{bg}}$ using 1,000 iterations of gradient descent with a learning rate that is exponentially decayed from 0.01 to 0.001, which takes less than four seconds on an RTX 4090 GPU in total.
In each iteration, we randomly sample $2^{21}$ viewing directions and backpropagate the mean squared error between sampling $\mathcal{F}_{\text{bg}}$ and the MLP onto the parameters of $\mathcal{F}_{\text{bg}}$.
Furthermore, we achieve a slight performance improvement during both training and inference by caching the local view directions, as these only depend on the intrinsic camera parameters and usually do not change between frames.
This is especially useful when the camera model includes lens distortion, as it avoids repeatedly undistorting the view directions for every frame.

\paragraph{Rendering Points as Small Gaussians.}
The bilinear splatting formulation used in INPC~\cite{hahlbohm2024inpc} has the advantage that, because the bilinear interpolation weights across the 2\texttimes2 fragments of each point sum to one, the total contribution of each point to the final image is the same regardless of its sub-pixel position.
However, the major issue with bilinear splatting is that the size of each point in screen space is the same regardless of distance to the camera and image resolution.
Naturally, this causes INPC to fail at rendering close-by objects accurately as the sampled point clouds are not dense enough leading to large holes which the CNN is unable to handle properly (see \cref{fig:teaser}).
Inspired by the success of 3DGS~\cite{kerbl3Dgaussians}, we propose to render points as small isotropic Gaussians during inference.
By default, we compute the world-space scale of these Gaussians so that the standard deviation of a Gaussian at the near plane that gets projected to the center of the image is no more than five pixels.
Similar to Kerbl et al.~\cite{kerbl3Dgaussians}, we use an affine approximation of the perspective projection~\cite{zwicker2001ewasplat} to project the Gaussian extent onto the image plane.
To tackle aliasing issues with distant points, we apply a smaller dilation to the projected 2D Gaussians to retain fine detail.
Specifically, the isotropic 2D Gaussian we use for dilation has a variance of $0.16$ -- Kerbl et al.\ use $0.3$ -- making it slightly larger than a pixel during rendering as we truncate Gaussians at $3\sigma$.

\subsection{Pre-Trained Hole-Filling CNN}\label{sec:pre_train}

Similar to other point-based neural rendering methods~\cite{aliev2020npbg, ruckert2022adop,kopanas2021perviewopt,franke2024trips}, INPC uses a CNN to decode the feature map output by the rasterizer while filling in any holes that may occur due the discrete nature of point clouds.
Usually, the weights of the CNN are randomly initialized as scenes are trained in isolation resulting in all model parameters being "overfitted" to a single scene.
For INPC, we observe that this significantly slows down convergence at the start of the training as early renderings are usually obscured due to the CNN being unable to decode the rasterizer output to a meaningful RGB output.
To ameliorate this problem, we develop a pre-training pipeline for INPC that results in a set of weights that can be used instead to initialize the parameters of the CNN.
For this work, we compose a small dataset for pre-training using all seven scenes from the \textit{Training Data} subset from Tanks and Temples~\cite{Knapitsch2017} as well as four scenes from CO3D~\cite{reizenstein2021co3d}.
While image resolution is roughly 2 megapixels in all scenes, the number of images varies significantly between 198 and 1,106.
We note that the selected pre-training scenes also contain photometric variations due to, e.g., auto-exposure being enabled during data capturing.

\paragraph{Stage 1: Ensuring a Common Latent Space.}
As a first step, we train a single INPC model on the \textit{Truck} scene from Tanks and Temples~\cite{Knapitsch2017} and store the final CNN weights to use them for initialization in the next stage.
While not specific to the \textit{Truck} scene, we found that this approach makes the information encoded in each of the four feature channels input into the CNN more coherent across different models.

\paragraph{Stage 2: Training INPC Models on a Pre-Training Dataset.}
In the next step, we train INPC models for all eleven pre-training scenes and initialize CNN weights using the output of the first stage.
To prevent the CNN from unlearning prior knowledge during early training iterations, we apply a cosine delay~\cite{barron2021mipnerf} to the learning rate of its parameters that reduces the learning rate during the first 12,500 iterations, i.e., $\nicefrac{1}{4}$ of the optimization.

\paragraph{Stage 3: Cross-Scene Pre-Training on Extracted Feature Maps.}
In the last step, we want to distill a single set of CNN weights from the models trained in the second stage.
To achieve this, we first extract the feature map outputs of the rasterizer for all viewpoints of all scenes.
Note that we would have to extract all parameters of the tonemapper in the INPC pipeline to enable proper supervision with the ground truth sRGB images.
As this would make the extraction and subsequent training significantly more complicated, we devise a modified, fully invertible implementation of the tonemapper instead.
While it is trivial to invert exposure correction ($I_{\text{e}} = \nicefrac{I_{HDR}}{2^{\text{EV}_i}},\ I_{HDR} = I_{\text{e}} \cdot 2^{\text{EV}_i}$), inverting the piecewise linear function approximating the camera response function is problematic as the implementation of Rückert et al.\cite{ruckert2022adop} only use a loss to encourages the endpoints, i.e., the first and last control point, to be 0 and 1 respectively.
Note how this makes accurate inversion impossible because the approximated camera response function is unable to represent the full range of possible output values.
We therefore actively enforce the endpoints of the approximate camera response function to be 0 and 1.
The resulting piecewise linear function now always covers the full range of possible outputs and is therefore also trivial to invert.
To maximize accuracy when passing the ground truth images to the inverted tonemapper, we perform all computations in double precision, which reduces the numerical error to 1e-16 in our tests.
We then use the extracted feature images and target HDR outputs to train a CNN with the same architecture, optimizer, and loss function used in the INPC pipeline.
During each of the 10,000 iterations, we select one random $512^2$ crop from each of the pre-trained scenes resulting in a batch size of 11.
To reduce overfitting, we extract crops with an original size randomly selected from a list of possible values $[512^2, 640^2, 768^2, 896^2, 1024^2]$, which are then bilinearly downscaled to $512^2$ to enable batched processing.
The learning rate is exponentially decayed from 6e-4 to 1e-4.

\paragraph{Usage of Pre-Trained Weights.}
When training with the pre-trained CNN weights resulting from the last stage, we also apply the cosine delayed learning rate schedule used in the second stage to prevent the CNN from unlearning prior knowledge.
We also found that the training in stage three is prone to overfitting, especially the FFC residual block.
As a consequence, we do not initialize the weights of the FFC residual block using the pre-trained weights, but use the default random initialization instead.

\begin{table*}[tb]
\centering
\setlength\tabcolsep{9.5pt}
\small
\begin{tabular}{@{}lccccc|ccccc}
\toprule
       & \multicolumn{5}{c|}{Mip-NeRF360~\cite{barron2022mipnerf360}} & \multicolumn{5}{c}{Tanks and Temples~\cite{Knapitsch2017}} \\
Method & PSNR$\uparrow$ & SSIM$\uparrow$ & LPIPS$\downarrow$ & Duration$\downarrow$ & VRAM$\downarrow$ & PSNR$\uparrow$ & SSIM$\uparrow$ & LPIPS$\downarrow$ & Duration & VRAM             \\ 
\midrule
INPC$^\dagger$     & 27.83 & 0.830 & 0.188 & \multicolumn{2}{c|}{--- not reported ---} & \cellcolor{1st}24.54 & 0.846 & 0.252 & \multicolumn{2}{c}{--- not reported ---} \\
INPC               & 27.83 & \cellcolor{1st}0.832 & \cellcolor{1st}0.185 & 4h49m & 19.2GiB                           & \multicolumn{5}{c}{--- out of memory ---}                        \\
Ours w$\slash$o PT & 27.89 & 0.830 & 0.186 & 3h57m & 16.7GiB                           & 24.25 & 0.846 & 0.251 & \cellcolor{1st}5h26m & 21.2GiB                         \\
Ours               & \cellcolor{1st}27.95 & \cellcolor{1st}0.832 & \cellcolor{1st}0.185 & \cellcolor{1st}3h55m & \cellcolor{1st}16.5GiB                           & \cellcolor{1st}24.54 & \cellcolor{1st}0.853 & \cellcolor{1st}0.239 & \cellcolor{1st}5h26m & \cellcolor{1st}21.1GiB                         \\
\bottomrule
\end{tabular}
\caption{%
Training performance on the Mip-NeRF360 and Tanks and Temples datasets. Our bag of tricks enables considerably faster training with similar or better quality. It also allows for training with higher resolution images, which was not possible with the original implementation. Numbers for the INPC version marked with $\dagger$ are copied from the original publication~\cite{hahlbohm2024inpc}. The best result is highlighted in \textcolor{1stText}{\textbf{green}}.
}\label{tab:training_comparisons}
\end{table*}

\subsection{Implementation Improvements}

The last category of tricks features multiple small performance and precision optimizations.

In INPC, a spherical contraction~\cite{barron2022mipnerf360} is applied to the sampled point positions before being input into $\mathcal{A}$, implemented in PyTorch.
This has a significant overhead because each computation is launched as a separate kernel.
We fuse the implementation into a single CUDA kernel to improve efficiency.
In a similar manner, we also fuse the customized Cauchy loss function based on the generalized robust loss function by J.~T.~Barron~\cite{barron2019general}.
In contrast to the spherical contraction, the Cauchy loss needs to be differentiable, which means we have to implement two kernels for the forward and backward pass respectively.
For the revised implementation, we also make use of the \verb|log1p| intrinsic to improve numerical precision.
As the multi-resolution hash grid used for $\mathcal{A}$ has around 300 million weights, computing the normalized weight decay as proposed by Barron et al.~\cite{barron2023ICCV} in PyTorch using automatic differentiation is costly in terms of both speed and VRAM usage.
Therefore, we opt to skip the loss computation during the forward pass and directly add the gradient of the loss to the gradient of each weight.
As we will show in our experiments, fusing these computations provides a surprisingly high speedup and a moderate reduction in VRAM usage.

\paragraph{Fused Rejection Sampling.}
To ensure unbiased sampling of the viewpoint-specific PDF, Hahlbohm et al.\ use a PyTorch-based rejection sampling approach during training.
This is necessary because random placement of points into voxels partially inside the viewing frustum can result in them being invisible during rendering.
Note how simply placing the point in the area of the voxel that is inside the frustum would introduce an unwanted bias toward these voxels.
Their PyTorch-based implementation therefore avoids this problem using multiple rounds of sampling in conjunction with tracking the ratio between visible and invisible samples in the first pass.
Subsequent passes then sample additional points based on this ratio until the desired number of visible samples is reached, which can lead to large spikes in VRAM usage.
To address this, we devise an improved implementation that fuses the rejection sampling process into a single CUDA kernel to fully eliminate excess VRAM usage.
Inside the kernel, we track the number of valid sample points created across all threads using an atomic counter.

\section{Experiments}

We evaluate our bag of tricks in a series of experiments.
Similar to Hahlbohm et al.~\cite{hahlbohm2024inpc}, we use a total of 17 scenes.
Nine scene from the Mip-NeRF360 dataset of which five are outdoor and four are indoor scenes, where image resolution is 1 and 1.5 megapixels respectively as well as eight scenes from the \textit{Intermediate} subset of the Tanks and Temples dataset~\cite{Knapitsch2017}, which have a resolution of around 2 megapixels.
Alongside training and rendering time as well as VRAM requirements, we report image quality using PSNR, SSIM, and LPIPS~\cite{zhang2018lpips}.
Our implementation is based on the original INPC implementation of Hahlbohm et al.~\cite{hahlbohm2024inpc} and is, unless otherwise noted in \cref{sec:tricks}, identical to theirs.
All experiments were conducted on a single RTX 4090 GPU.

\subsection{Training Improvements}

Many of our proposed modifications aim at improving efficiency during training.
We show their joint impact in \cref{tab:training_comparisons}.
In comparison to the original INPC~\cite{hahlbohm2024inpc}, we achieve a sizable reduction in training time and VRAM requirements.
We highlight that the reduced VRAM requirements enable our method to train the higher resolution scenes from the Tanks and Temples dataset~\cite{Knapitsch2017} on consumer grade GPUs -- something that is not possible with the original implementation.

\begin{figure}
\centering
\includegraphics[width=.47\textwidth]{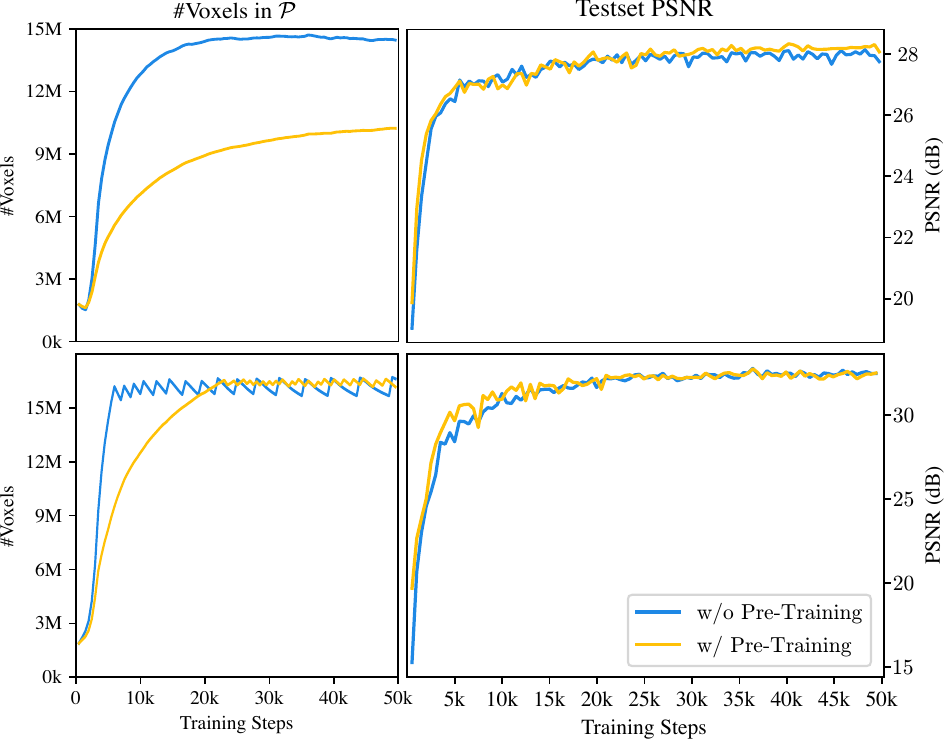}
\caption[short]{%
Convergence plots showing the number of voxels in $\mathcal{P}$ and PSNR on the testset for the \textit{Bicycle} (top) and \textit{Bonsai} (bottom) scenes from Mip-NeRF360~\cite{barron2022mipnerf360} over the duration of training. The reduction in active voxels indicates more efficient empty space pruning, which we find to increase robustness of the optimization. We consistently observed this effect across all tested scenes.
}\label{fig:convergence_plots}
\end{figure}

Our pre-training strategy, detailed in \cref{sec:pre_train}, marginally improves image quality upon full convergence, as shown in \cref{tab:training_comparisons}. Early training iterations produce higher quality renderings (\cref{fig:convergence_plots} (right)), aiding in avoiding local minima and preventing incorrect voxel pruning in the probability field. Furthermore, using pre-trained CNN weights consistently lowers active voxels during optimization across all scenes, as depicted in \cref{fig:convergence_plots} (left).
Consequently, our method delivers superior quality within a restricted time frame (e.g., 5,000 iterations) and produces a more uniform voxel distribution, presenting an interesting avenue for future research.

\begin{table}[tb]
\centering
\setlength\tabcolsep{12pt}
\small
\begin{tabular}{@{}lcc}
\toprule
Modification                & $\Delta\nicefrac{\text{Time}}{\text{Iteration}}$ & $\Delta$ VRAM Peak \\ 
\midrule
Tiled Rasterizer            & - 17.77 ms     & - 0.37 GiB \\
Fused Weight Decay          & - 45.46 ms     & - 2.18 GiB \\
Fused Rejection Sampling    & - 1.72 ms      & - 3.20 GiB \\
Fused Spherical Contraction & - 1.03 ms      & - 0.22 GiB \\
Fused Cauchy Loss           & - 0.12 ms      & - 0.05 GiB \\
Cached Pixel Directions     & - 0.08 ms      & $\pm$ 0.0 GiB \\
\bottomrule
\end{tabular}
\caption{%
The impact of each modification that influences the duration of a single training iterations and as well as peak VRAM usage during training on the \textit{Bicycle} scene from Mip-NeRF360~\cite{barron2022mipnerf360}. For the fused rejection sampling, the reported VRAM delta corresponds to the worst-case overhead of the original implementation.
}\label{tab:training_ablations}
\end{table}

In \cref{tab:training_ablations}, we isolate the contribution of each modification with respect to reducing training time and peak VRAM usage.
The fused weight decay implementation has the most significant impact on training time followed by our improved rasterizer implementation.
The fused rejection sampling helps to avoid large peaks in VRAM usage that occur in the original implementation when a lot of voxels are close to the camera and partially inside the viewing frustum.

\subsection{Inference Improvements}

Multiple of the proposed modifications have an impact on rendering time during inference, e.g., the fused spherical contraction (see \cref{tab:training_ablations}).
We observe the distilled background model to reduce time spent evaluating it from between one and two milliseconds to a few microseconds depending on the image resolution.
However, the major improvements for inference are due to our revised sampling techniques and the tiled renderer supporting rendering of points as small Gaussians.

\paragraph{View-Specific Inference Rendering.}
\begin{table}[tb]
\centering
\setlength\tabcolsep{8.6pt}
\small
\begin{tabular}{@{}lcccc}
\toprule
Method & PSNR$\uparrow$ & SSIM$\uparrow$ & LPIPS$\downarrow$ & Render$\downarrow$ \\ 
\midrule
INPC$^\dagger$             & 27.83 & \cellcolor{3rd}0.830 & 0.188 & ---     \\
INPC                       & 27.83 & \cellcolor{1st}0.832 & \cellcolor{2nd}0.185 & 114.2ms \\
Ours                       & \cellcolor{1st}27.95 & \cellcolor{1st}0.832 & \cellcolor{2nd}0.185 & \cellcolor{2nd}93.9ms  \\
Ours w$\slash$ Gaussians   & \cellcolor{3rd}27.87 & \cellcolor{1st}0.832 & \cellcolor{1st}0.183 & \cellcolor{3rd}96.4ms  \\
Ours w$\slash$ ring buffer & \cellcolor{2nd}27.88 & \cellcolor{2nd}0.831 & \cellcolor{3rd}0.186 & \cellcolor{1st}46.4ms     \\
\bottomrule
\end{tabular}
\caption{%
Inference performance with view-specific multi-sampling on the nine scenes from the Mip-NeRF360 dataset. Numbers for the INPC version marked with $\dagger$ are copied from the original publication~\cite{hahlbohm2024inpc}. The three best results are highlighted in \textcolor{1stText}{\textbf{green}} in descending order of saturation.
}\label{tab:multisampled_inference}
\end{table}

In \cref{tab:multisampled_inference}, we show the results of different model configurations in comparison to the original INPC~\cite{hahlbohm2024inpc}.
Our ring buffer-based approach for reusing point clouds of previous frames has a significant impact on rendering speed, reducing the time it takes to compute each frame by more than 2\texttimes.
This makes rendering implicit neural point clouds a lot faster without sacrificing quality and even makes rendering more temporally stable as the geometry changes between frames are much smaller.
Using our improved tiled rasterizer with the bilinear splatting formulation provides the highest quality and is slightly faster than the original implementation.
Rendering points as small Gaussians, only slightly impacts quality metrics but makes a major difference when exploring a scene in a graphical user interface -- especially for close-up views.
The effect can be seen in \cref{fig:teaser} (left).

\paragraph{Inference with a Pre-Extracted Point Cloud.}
\begin{table}[tb]
\centering
\setlength\tabcolsep{7.5pt}
\small
\begin{tabular}{@{}lcccc}
\toprule
Method & PSNR$\uparrow$ & SSIM$\uparrow$ & LPIPS$\downarrow$ & Render$\downarrow$ \\ 
\midrule
3DGS~\cite{kerbl3Dgaussians} & \cellcolor{3rd}27.20 & \cellcolor{3rd}0.814 & 0.254 & \cellcolor{1st}3.2ms \bstrut \\
\midrule
INPC$^\dagger$               & 26.85 & 0.802 & 0.207 & ---   \tstrut \\
INPC                         & 26.85 & 0.803 & \cellcolor{3rd}0.205 & 37.3ms        \\
Ours (INPC sampling)         & 26.96 & 0.802 & \cellcolor{2nd}0.203 & \cellcolor{3rd}22.8ms        \\
Ours                         & \cellcolor{1st}27.47 & \cellcolor{1st}0.824 & \cellcolor{1st}0.184 & \cellcolor{2nd}22.6ms        \\
Ours w$\slash$ Gaussians     & \cellcolor{2nd}27.35 & \cellcolor{2nd}0.823 & \cellcolor{1st}0.184 & 25.0ms        \\
\bottomrule
\end{tabular}
\caption{%
Inference performance with a pre-extracted global point cloud on the nine scenes from the Mip-NeRF360 dataset. INPC uses 67M points, while our revised pre-extraction works better with 33M points. Numbers for the INPC version marked with $\dagger$ are copied from the original publication~\cite{hahlbohm2024inpc}. As a benchmark, we also include results of 3D Gaussian Splatting. The three best results are highlighted in \textcolor{1stText}{\textbf{green}} in descending order of saturation.
}\label{tab:preextracted_inference}
\end{table}

We show the impact of our improved algorithm for global point cloud extraction in \cref{tab:preextracted_inference}.
Similar to rendering view-specific point clouds, our tiled rasterizer provides a major speedup that leads to real-time frame rates.

We further observe slightly improved quality over INPC~\cite{hahlbohm2024inpc} when using their extraction algorithm with our models.
We highlight that our proposed algorithm leads to a major improvement in quality metrics, while only requiring half the number of points (33M).
This improvement causes our method to consistently outperform 3DGS~\cite{kerbl3Dgaussians} in terms of quality.
Due to the different distribution of points, however, we observe no meaningful rendering speedup compared to using the larger point clouds produced by INPC's algorithm (67M points).
Still, we find that the point clouds obtained using our revised algorithm are of higher quality and model especially the background with higher fidelity.
Rendering points as small Gaussians again improves visual fidelity in close-up views as evident in \cref{fig:teaser} (right).

\section{Open Challenges}

Despite the efficiency gains of our importance‐driven sampling framework, several challenges remain.
First, the lack of true volumetric rendering prevents the removal of view-specific multisampling, which remains expensive even when applying our ring buffer approach.
Second, view‐dependent effects modeled by our hole-filling CNN exhibit temporal instability, suggesting the need for architectures or loss terms that explicitly enforce coherence over time.
Third, as depth sorting still dominates the runtime in complex scenes, hybrid order-independent transparency schemes~\cite{hahlbohm2025htgs} may offer a practical compromise.
Fourth, we found that changes to the appearance model, e.g., having it predict 3D Gaussian or beta-primitives~\cite{liu2025betasplatting}, disrupts sampling-weight convergence during optimization, indicating that more robust, renderer‑agnostic update rules are needed.
Fifth, in large, multi-room environments the view-specific sampling algorithm wastes effort on occluded regions; incorporating, e.g., occlusion rays could focus computation on what is actually visible~\cite{Fricke2024}.
Finally, probability fields exceeding $256^3$ active voxels suffer floating‐point precision errors near the camera; hierarchical or mixed‑precision representations might preserve accuracy without blowing up memory.
Tackling these issues will be crucial for bringing real‑time, high‑fidelity rendering to ever more complex and dynamic scenes.

\section{Conclusion}
We have presented a set of targeted improvements to the INPC pipeline that significantly enhance its computational performance without compromising image quality. 
Our optimizations -- including improved sampling, a faster rasterizer, and pre-training of the hole-filling CNN -- enable more interactive rendering and scalable deployment. 
These enhancements are modular and general enough to benefit similar hybrid neural rendering methods.

Our experiments demonstrate that applying our bag of tricks to INPC doubles rendering speed, achieves up to a 25\% decrease in training time, and lowers VRAM usage, making INPC a viable alternative to explicit representations like 3D Gaussian Splats, especially on resource-constrained devices.

\section*{Acknowledgments}
We would like to thank Timon Scholz for his help with the evaluation, as well as Brent Zoomers and Moritz Kappel for their valuable suggestions and early discussions.
This work was partially funded by the DFG (“Real-Action VR”, ID 523421583) and the L3S Research Center, Hanover, Germany.

\bibliographystyle{eg-alpha-doi} 
\bibliography{references}

\appendix
\section{Further Improvements and Modifications}

The most important changes our paper contributes are already described under \Cref{sec:tricks}. However, in the following we also report and comment on other modifications we implemented over the original code for INPC as proposed by Hahlbohm and colleagues~\cite{hahlbohm2024inpc}.

\paragraph{Addressing Gradient Approximations.}

The computation of depth-sorted alpha blending is governed by:
\begin{equation}
    \label{eq:alpha_blending}
	\mathcal{F}_{\text{pixel}} = \sum_{i=1}^N T_i \alpha_i f_i \hspace{1.5em} \text{and} \hspace{1.5em}
	T_i = \prod_{j=1}^{i-1}(1-\alpha_j),
\end{equation}
where $\alpha_i$ indicates the opacity of fragment $i$.

Note how the partial derivative of \cref{eq:alpha_blending} with respect to a $k$-th fragment that has $\alpha_k=0$ is
\begin{equation}
\frac{\partial\mathcal{F}_{\text{pixel}}}{\partial\alpha_k} = T_k f_k - \sum_{i=k+1}^K \alpha_i T_i f_i\ +\ T_{K+1} f_{\text{bg}}.
\end{equation}
We observed that the implementation of Hahlbohm et al.\ skips these computations which may lead to unexpected side-effects during gradient descent.
When we investigated this, we also found the same issue in the source code of the popular 3DGS implementation by Kerbl et al.~\cite{kerbl3Dgaussians} and suggest that it would be interesting to investigate its impact.
Our rudimentary tests showed slightly improved background reconstruction but also a minor increase in floaters on some scenes.

Furthermore, Hahlbohm et al.\ discuss in their source code that they got slightly better results with an accidentally introduced mistake in the feature gradient of each point.
For improved extensibility and reliability, we correct both issues in our implementation.

\paragraph{Additional Modifications} %
INPC uses GELU activations~\cite{hendrycks2016gelu} inside the hole-filling CNN.
We replace them with SiLU activations as they are slightly faster and support in-place computation in PyTorch.
Lastly, we adjust the learning rate used for the camera response function of $\mathcal{T}$ to 1e-4 instead of 1e-3 to allow for better fitting.

\end{document}